\documentclass[a4paper,fleqn]{cas-dc}

\usepackage[authoryear,longnamesfirst]{natbib}

\def\tsc#1{\csdef{#1}{\textsc{\lowercase{#1}}\xspace}}
\tsc{WGM}
\tsc{QE}
\tsc{EP}
\tsc{PMS}
\tsc{BEC}
\tsc{DE}

\begin{document}
\let\WriteBookmarks\relax
\def\floatpagepagefraction{1}
\def\textpagefraction{.001}

\shorttitle{Patient-Made Knowledge Networks: Long COVID
Discourse, Epistemic Injustice, and Online Community
Formation}

\shortauthors{Ammari}

\title [mode = title]{Patient-Made Knowledge Networks: Long COVID
Discourse, Epistemic Injustice, and Online Community
Formation}                      



%
\author[1]{Tawfiq Ammari}[
                        orcid=0000-0002-1920-1625]
\cormark[1]
\ead{tawfiq.ammari@rutgers.edu}

\affiliation[1]{organization={Rutgers School of Communication and Information},
    city={New Brunswick},
    state={NJ},
    country={USA}}

\cortext[cor1]{Corresponding author}



\begin{abstract}
Long COVID represents an unprecedented case of patient-led illness definition, emerging through Twitter in May 2020 when patients began collectively naming, documenting, and legitimizing their condition before medical institutions recognized it. This study examines 2.8 million tweets containing \#LongCOVID to understand how contested illness communities construct knowledge networks and respond to epistemic injustice. Through topic modeling, reflexive thematic analysis, and exponential random graph modeling (ERGM), we identify seven discourse themes spanning symptom documentation, medical dismissal, cross-illness solidarity, and policy advocacy. Our analysis reveals a differentiated ecosystem of user roles—including patient advocates, research coordinators, and citizen scientists—who collectively challenge medical gatekeeping while building connections to established ME/CFS advocacy networks. ERGM results demonstrate that tie formation centers on epistemic practices: users discussing knowledge sharing and community building formed significantly more network connections than those focused on policy debates, supporting characterization of this space as an epistemic community. Long COVID patients experienced medical gaslighting patterns documented across contested illnesses, yet achieved WHO recognition within months—contrasting sharply with decades-long struggles of similar conditions. These findings illuminate how social media affordances enable marginalized patient populations to rapidly construct alternative knowledge systems, form cross-illness coalitions, and contest traditional medical authority structures.

\end{abstract}


\begin{highlights}
\item Long COVID emerged as the first patient-defined illness condition through social media organizing
\item Analysis of 2.8 million tweets reveals seven discourse themes spanning symptoms, epistemic injustice, and policy advocacy
\item ERGM analysis shows network ties form around epistemic practices rather than policy debates
\item Twitter enabled rapid connection to ME/CFS advocacy networks, accelerating knowledge transfer across contested illnesses
\item Long COVID achieved WHO recognition in months, contrasting decades-long struggles of similar conditions
\end{highlights}


\maketitle

\section{Introduction}

Long COVID represents a critical case for understanding how contested illnesses emerge and gain legitimacy in contemporary medicine. Defined by the WHO as symptoms persisting beyond three months after initial SARS-CoV-2 infection \citep{who2023}, long COVID affects an estimated 17 million people in Europe alone \citep{turner2023longcovid}. The condition emerged not through traditional medical channels, but through patient self-organization on Twitter in May 2020, when Dr. Elisa Perego first used the hashtag \#longcovid to describe her ``cyclical, progressive, and multiphasic'' symptoms \citep{callard_how_2021}.

The emergence of long COVID through social media highlights fundamental tensions in medical knowledge production. As Callard and Perego \citeyearpar{callard_how_2021} demonstrate, long COVID is the first illness condition created entirely through patients finding each other online, bypassing traditional medical gatekeepers. This patient-led construction of illness categories challenges the authority structures that typically govern medical knowledge \citep{turner2023longcovid}.

Long COVID patients confronted epistemic injustice—the systematic discrediting of patient knowledge by medical authorities \citep{fricker_epistemic_2007}. Similar dynamics have been documented in other contested illnesses like ME/CFS, fibromyalgia, and Lyme disease, where predominantly female patient populations face medical dismissal and psychologization \citep{nezamdoust2025contested, morgan1998contested}. However, the rapid emergence of long COVID during a global pandemic, combined with unprecedented use of social media for patient organizing, created novel dynamics in how contested illnesses can gain recognition.

This study examines how long COVID discourse evolved on Twitter from May 2020 to July 2022, focusing on three key questions:

\begin{enumerate}
\item How did long COVID patients frame their illness experience and challenge medical authority?
\item What roles emerged within the long COVID networked public, and how did network structure and tie formation patterns reflect the constitution of an epistemic community?
\item How does the long COVID case illuminate broader patterns of epistemic injustice in contested illness?
\end{enumerate}

By analyzing 2.8 million tweets through topic modeling and network analysis, we demonstrate how marginalized patient communities leverage digital platforms to construct collective knowledge, challenge medical gatekeeping, and advocate for policy change. This contributes to medical sociology's understanding of patient activism, contested illness, and the changing dynamics of medical authority in the digital age.

\section{Background}

\subsection{Contested Illness and Epistemic Injustice}

Contested illnesses exist at the margins of medical legitimacy, lacking clear biomarkers, definitive diagnoses, and effective treatments \citep{conrad2010social, dumit2006illnesses}. Conditions like ME/CFS, fibromyalgia, and chronic Lyme disease are contested because their symptoms remain medically suspect, leading to disputes between patients and medical authorities over the ``realness'' of disease \citep{barker2011listening, nezamdoust2025contested}.

Foucault \citeyearpar{foucault1975birth} argues that medical professionals define normalcy through standardized training, positioning them as central arbiters of what constitutes legitimate illness. When patients present symptoms that do not fit standardized disease categories, they face testimonial injustice—their illness narratives are systematically discredited \citep{fricker_epistemic_2007}. This is compounded by hermeneutical injustice: the lack of shared language to describe experiences that fall outside established medical frameworks \citep{fricker_epistemic_2007}.

Gender intersects critically with contested illness. Approximately 80\% of ME/CFS patients are women \citep{meresearch2015}, reflecting broader patterns of medical sexism where women's symptoms are more likely to be dismissed as psychosomatic \citep{khan2024psychological}. Nezamdoust and Ruel \citeyearpar{nezamdoust2025contested} demonstrate how severe ME/CFS patients face both functional debilitation and social invisibility, excluded from the very disability systems meant to support them. Åsbring and Närvänen \citeyearpar{aasbring2003ideal} found that physicians treating CFS and fibromyalgia patients experienced frustration due to the uncertainty inherent in these illnesses, often categorizing them as ``illnesses" rather than ``diseases" due to lack of objective measurable values. This medical skepticism directly shapes patient experiences, as physicians struggle between their professional ideals and the reality of treating contested conditions. The pattern of gendered medical marginalization appears across contested illnesses \citep{goudsmit2009learning}.

The psychologization of contested illness intensifies stigma by implicitly blaming patients for their condition. Cognitive Behavioral Therapy (CBT) and Graded Exercise Therapy (GET), long recommended for ME/CFS despite evidence of harm, frame symptoms as resulting from ``dysfunctional thoughts and beliefs'' \citep{kindlon2011reporting, vink2022updated}. This therapeutic approach ignores physical limitations like post-exertional malaise while positioning patients as responsible for their own suffering.

Diagnosis serves not merely as a label but as social validation \citep{jutel2009sociology, jutel2010medically}. Without legitimate diagnoses, patients lose access to the ``sick role'' \citep{Parsons1951-PARTSS-7}—the socially sanctioned exemption from normal obligations. They face stigmatization for failing to be productive members of society \citep{dumit2006illnesses, nettleton2006just}. For contested illness patients, the struggle for diagnosis becomes a fight for social legitimacy and basic recognition of suffering.

\subsection{Online Patient Communities and Health Activism}

Social media has transformed how patients with stigmatized conditions find community and voice their experiences. Online health communities enable patients to share illness trajectories, compare symptoms, and build collective expertise while maintaining anonymity that reduces stigma \citep{ammari_networked_2015, macleod_rare_2015, sannon_i_2019}.

Wenger's \citeyearpar{wenger1999communities} concept of communities of practice illuminates how chronic disease patients collectively learn and develop shared knowledge systems. Online health communities create spaces where ``collective learning takes place,'' allowing patients to ``constitute experiential knowledge and build collective expertise'' \citep{akrich2010communities}. For example, diabetic patient communities collaboratively reassembled illness trajectories by sharing data about blood sugar responses to different interventions \citep{huh_collaborative_2012}.

Some online patient communities evolve from communities of practice into epistemic communities—groups that share policy orientations and work to shape how illness is understood and treated \citep{haas1992introduction}. Mankoff et al. \citeyearpar{mankoff_disability_2010} documented how a Lyme disease community developed an alternative disease model that better fit their lived experiences, then advocated for laws protecting physicians who treated according to this minority model.

Brown et al. \citeyearpar{brown2004embodied} theorized embodied health movements (EHMs) as a distinct category of health social movements that: (1) center the biological body and embodied illness experience, (2) challenge existing medical/scientific knowledge and practice, and (3) involve activists collaborating with scientists and health professionals. Zavestoski et al. \citeyearpar{zavestoski2004patient} demonstrated how Gulf War illness patients, like those with CFS, fibromyalgia, and multiple chemical sensitivity, struggled for diagnostic legitimacy amid medical uncertainty. They identified four factors in diagnostic legitimacy: medical community recognition, lay acceptance, uncertainty about causes, and social mobilization. Their work shows that ``medically unexplained physical symptoms" become sites of contestation where patient activism directly challenges medical authority \citep{zavestoski2004patient}.

Twitter offers unique affordances for health activism through its public, hashtag-based structure that enables rapid community formation and wide message dissemination \citep{xu2015twitter}. For long COVID specifically, Turner et al. \citeyearpar{turner2023longcovid} demonstrate how Twitter enabled patients to establish social consensus around their varied symptoms, overcoming traditional evidence-based medicine's difficulty capturing intermittent and multisystem presentations. The hashtag \#researchrehabrecognition became a rallying cry for medical legitimacy.

Long COVID patients benefited from pandemic-era attention to COVID-19 and the sheer number of affected individuals, enabling faster recognition than previous contested illness movements \citep{callard_how_2021, turner2023longcovid}.

\subsection{Patient Agency and Social Support in Chronic Illness}

Beyond community formation, individual patient experiences of contested illness are shaped by agency—their capacity to take action in managing health—and social support from family, friends, and healthcare providers. Figueiredo et al. \citeyearpar{figueiredo2024journey} developed a framework showing how these factors interact to shape long COVID trajectories.

Patients with both high agency and strong social support actively engaged in self-management, experimented with coping strategies, and advocated effectively within healthcare systems. In contrast, those with low agency and limited support faced compounded challenges: difficulty navigating healthcare, limited ability to self-advocate, and reduced access to information and resources \citep{figueiredo2024journey}. Medical gaslighting and dismissal further undermined patient agency, creating cycles of disempowerment.

This framework helps explain variation in how contested illness patients navigate their conditions. Social support networks can either reinforce medical legitimacy or perpetuate stigma. Healthcare providers who take patients seriously can enhance agency, while those who dismiss symptoms erode it \citep{figueiredo2024journey, nezamdoust2025contested}.

\section{Methods}

This study employs a mixed-methods approach combining computational analysis with qualitative interpretation, informed by critical discourse analysis. Following Turner et al.'s \citeyearpar{turner2023longcovid} reflexive approach to analyzing long COVID Twitter discourse, we use computational methods to identify patterns in large-scale data while grounding interpretation in sociological theory about contested illness, epistemic injustice, and patient activism.

\begin{figure*}
    \centering
    \includegraphics[width=1\linewidth]{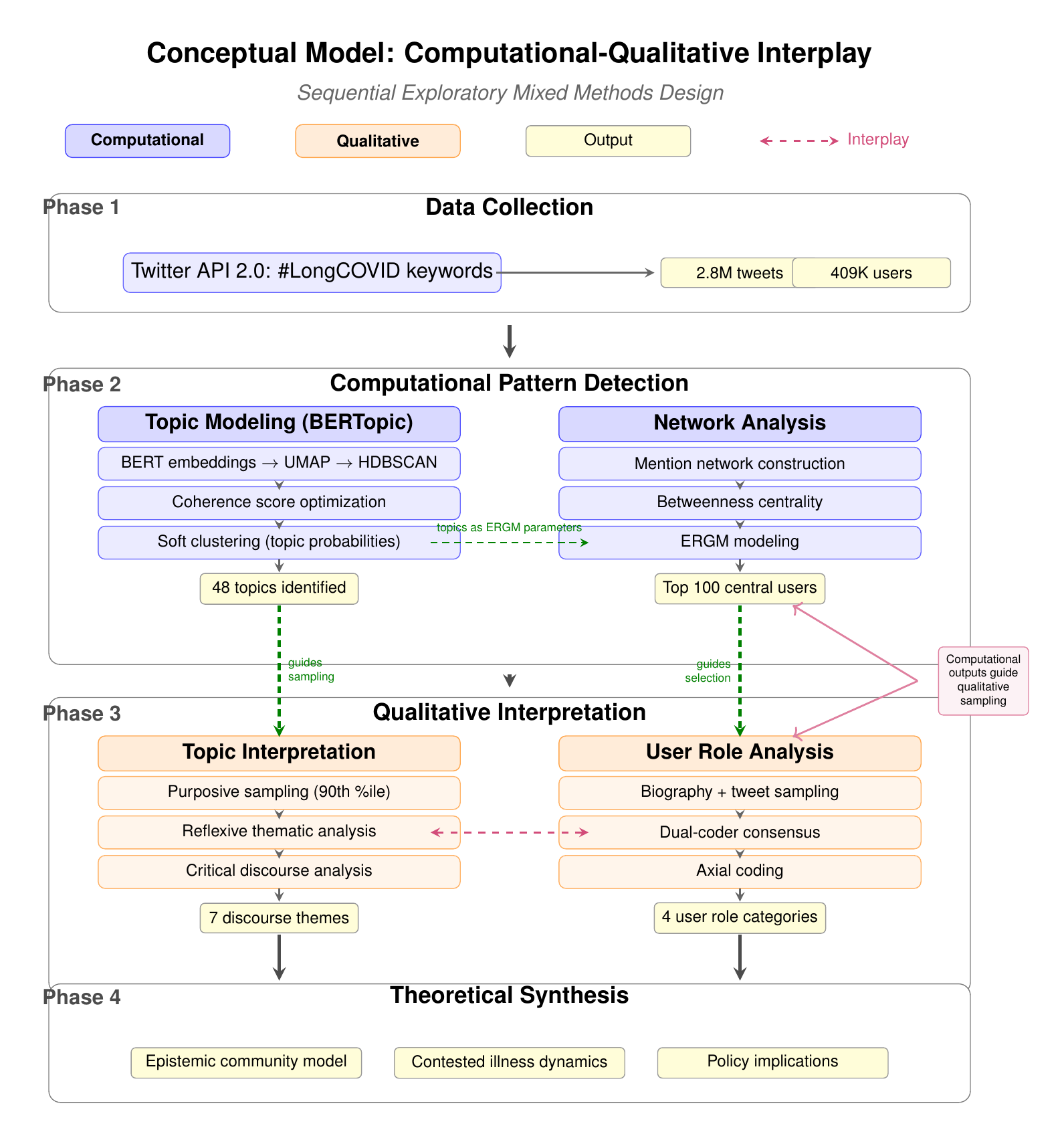}
    \caption{Conceptual model of computational-qualitative interplay across four analytical phases. Blue boxes indicate computational methods; orange boxes indicate qualitative methods; yellow boxes indicate outputs. Green dashed arrows show how computational outputs guide qualitative sampling ("scaffolding"). Purple dashed arrows indicate bidirectional interplay between parallel analytical streams. The model demonstrates sequential exploratory design where computational pattern detection identifies what exists in the data, while qualitative interpretation explains why patterns matter sociologically.}
    \label{fig:conceptualmeth}
\end{figure*}

This study employs a mixed-methods design following a sequential exploratory strategy (Creswell \& Creswell, 2017). The initial phase used computational methods—topic modeling and network analysis—to identify discourse patterns and structural positions within the long COVID Twitter network. These computational findings then guided purposive sampling for qualitative analysis, where reflexive thematic analysis (Braun \& Clarke, 2006) was applied to interpret user roles and epistemic practices. This approach aligns with emerging methodological frameworks in computational social science that position quantitative pattern detection as scaffolding for deeper qualitative interpretation \citep{nelson2020computational}, rather than as ends in themselves. The computational methods reveal what patterns exist in the data; the qualitative analysis explains why these patterns matter sociologically and how they illuminate processes of epistemic community formation and contested illness discourse. Figure \ref{fig:conceptualmeth} summarizes this study's analytical pipeline.

\subsection{Data Collection}

Using the Twitter API 2.0, we collected all tweets containing the hashtag \#LongCovid (and variants \#longcovid, \#LongCOVID), as well as tweets containing the phrases ``long covid,'' ``longcovid,'' ``long haul,'' or ``long haulers.'' Data collection spanned from the first use of the hashtag on May 20, 2020 to June 30, 2022, capturing the emergence and establishment of long COVID as a recognized condition---from Perego's initial tweet through WHO's formal definition \citep{callard_how_2021, turner2023longcovid}. This yielded 2,818,709 tweets from 409,272 unique users. Of these, 2,358,610 were retweets.

\subsection{Topic Modeling: BERTopic}

Topic modeling is an unsupervised machine learning technique that identifies latent thematic structures within a text corpus. Traditional approaches like Latent Dirichlet Allocation (LDA) use bag-of-words representations that fail to capture semantic context, performing poorly on short texts like tweets \citep{jonsson2015evaluation}. We employed BERTopic \citep{grootendorst2022bertopic}, which uses BERT to produce contextualized embeddings that capture meaningful semantic relationships—recognizing, for example, that ``fatigue" and ``exhaustion" are closer to each other than to ``policy." To create topic clusters, we reduced the high-dimensional embeddings using UMAP, then clustered contextually similar documents using HDBSCAN, a density-based algorithm that can identify outliers rather than forcing all documents into clusters \citep{grootendorst2022bertopic}

\subsubsection{Document Definition and Model Selection}

We defined each document as the aggregated text of all tweets by a single user, excluding retweets to reduce bias \citep{schofield2017understanding}, yielding 65,285 documents. To determine the optimal number of topics, we trained 25 models by varying HDBSCAN's minimum cluster size parameter (15–255, increments of 10) and evaluated them using coherence scores, which better approximate human judgments of topic interpretability than metrics like perplexity \citep{roder_exploring_2015}. The model with the highest coherence score (0.581; minimum cluster size = 115) produced 47 topics. We used HDBSCAN's soft-clustering technique to assign each tweet a probability vector indicating its relationship to multiple topics, capturing how topics overlap \citep{grootendorst2022bertopic}.

\subsection{Qualitative Analysis of Topics}

Topic models require human interpretation to become meaningful \citep{dou2011paralleltopics}. Our qualitative analysis proceeded in three stages:

\subsubsection{Sampling Strategy}

For each of the 47 topics, we randomly sampled 100 tweets where the topic probability score fell in the top 90th percentile, yielding an initial sample of 4,700 tweets. When initial samples proved insufficient for determining topic meaning, we resampled additional tweets. Some topics required resampling because we could not reach consensus on the themes using the first sample, ultimately yielding 5,100 tweets for analysis.

\subsubsection{Coding Process and Consensus Building}
\label{qualthemecoding}
To ensure methodological rigor, we employed intercoder dependability \citep{o2020intercoder, cascio2019team}. I coded sampled tweets for each topic until reaching saturation, then assigned descriptive labels. A second coder independently analyzed the same materials, with weekly meetings to compare codes and resolve discrepancies through consensus. This process led to refinements—for instance, separating ``research coordinator" from ``information curator" based on whether users primarily recruited participants or synthesized findings. In a second iteration, we used axial coding to identify relationships across topics and "systematically develop categories" \citep{strauss1990basics}, yielding seven thematic clusters from the original 47 topics.

\subsubsection{Critical Discourse Analysis}

We grounded our interpretation in critical discourse analysis (CDA), which examines how language reflects and reproduces social power relations \citep{fairclough2023critical}. Fairclough conceptualizes discourse as a three-dimensional phenomenon linking texts, discursive practices (production and consumption), and wider social structures. CDA enabled us to analyze how long COVID patients' tweets both responded to and challenged dominant medical discourses that delegitimized their experiences.

Specifically, we attended to how patients constructed counter-narratives to medical authority, how they claimed epistemic credibility for experiential knowledge, and how collective framing shifted individual suffering into political demands. This analytical lens connected micro-level tweet content to macro-level structures of medical authority and epistemic injustice documented in the contested illness literature \citep{fricker_epistemic_2007,barker2011listening, dumit2006illnesses}.

\subsection{Network Analysis}

Twitter users interact through multiple mechanisms: following accounts, retweeting posts, and mentioning other users. While retweet networks capture message amplification \citep{gallagher2021sustained}, mention networks better identify ``central and influential users'' through direct engagement \citep{auxier_handsoffmyada_2019}. I constructed a directed mention network where a link from user \textit{u} to user \textit{v} indicates that \textit{u} mentioned \textit{v} in a tweet. The resulting network contained 61,178 nodes and 317,594 edges.

To identify central actors, I calculated betweenness centrality---the extent to which a node lies on paths between other nodes in the network \citep{newman_networks_2018}. Users with high betweenness serve as information bridges, connecting different subgroups within the community. This measure identifies not necessarily the most followed accounts, but those who facilitate information flow across otherwise disconnected clusters.

\subsubsection{Exponential Random Graph Models}

To understand how network structure emerges, I employed Exponential Random Graph Models (ERGM), which extend multivariate regression to relational data: the presence or absence of a tie is predicted by structural and attribute-based parameters \citep{robins_lusher_2012}. Unlike descriptive network metrics, ERGM explains how endogenous and exogenous actor-level attributes affect tie formation \citep{keegan_et_al_2012}.

Using Statnet \citep{handcock2019package}, I modeled the mentions network with endogenous parameters (edges as baseline density; mutuality for reciprocity) and exogenous parameters (verified account status; sentiment valence via VADER \citep{hutto2014vader}; and discussion themes identified through topic modeling in \S\ref{qualthemecoding}). Positive parameters increase tie likelihood; negative parameters decrease it.

\subsection{Qualitative Coding of Central Users}

We qualitatively coded the 100 users with highest betweenness centrality to understand what roles emerged within the long COVID networked public. Following Pathak et al. \citeyearpar{pathak2021method}, who demonstrate that Twitter biographies contain adequate linguistic signifiers for identifying social identities users wish to portray, we analyzed each user's biography field alongside a random sample of 10 tweets per user.

When role determination proved difficult, we resampled additional tweets from that user's timeline---this occurred for three users. Importantly, because our dataset only captured tweets containing long COVID hashtags or keywords, we also sampled from users' broader timelines to understand their fuller identity and engagement patterns.

User roles were not mutually exclusive; a single account could occupy multiple positions. For example, a researcher might also identify as an advocate calling for more funding. Two coders independently analyzed all 100 accounts. We compared codes and reached consensus through discussion when classifications diverged \citep{macdonal_schoenebeck_19}. This process yielded four main categories: Patient Advocates and Lived Experience Experts; Medical and Research Professionals; Knowledge and Information Roles; and Policy and Advocacy.

\subsection{Ethical Considerations}

Following AoIR Ethical Guidelines 3.0, which caution against sharing datasets containing sensitive information that "could be used directly or indirectly against individuals" \citep[p.45]{franzke2020internet}, we do not release our dataset. Consistent with Fiesler and Proferes' \citeyearpar{fiesler_participant_2018} finding that Twitter users generally do not expect verbatim quotation in research, we mostly describe tweet content rather than quote directly. When direct quotes appear, we use only organizational accounts (e.g., @LongCovidSOS, @SurvivorCorps) or apply Bruckman's \citeyearpar{bruckman_studying_2002} recommended levels of disguise to protect user privacy. 

\section{Findings}

\begin{figure*}
    \centering
    \includegraphics[width=1\linewidth]{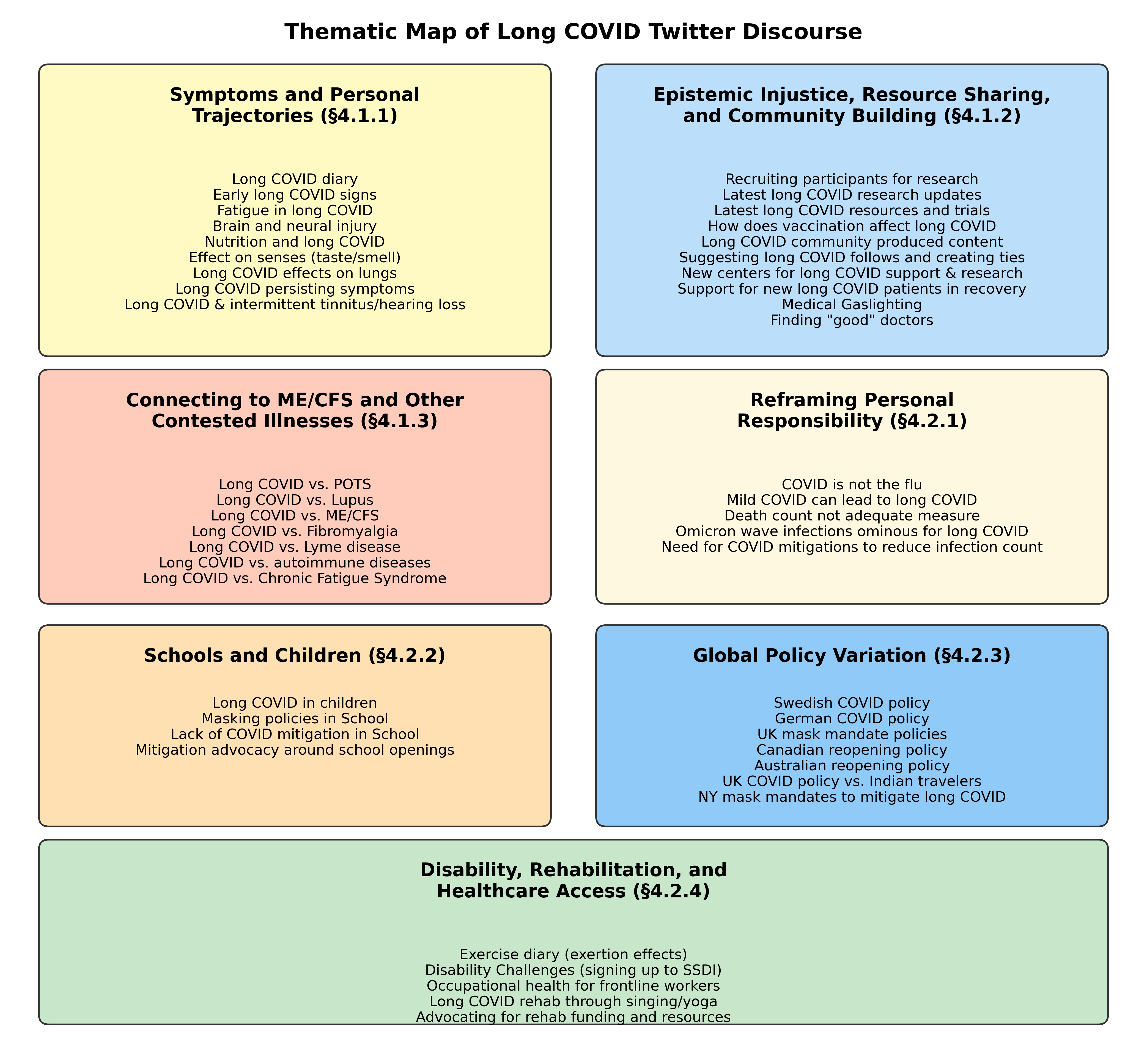}
    \caption{Thematic map of long COVID discourse on Twitter, showing seven themes and associated topics derived from reflexive thematic analysis of user-generated content.}
    \label{fig:thememap}
\end{figure*}

Our reflexive thematic analysis generated seven themes that reveal how patients navigate the epistemic uncertainties of a novel, contested condition through collective sense-making on social media (see Figure \ref{fig:thememap}). We expand on each of these themes in our findings.

\subsection{Framing Long COVID: From Personal Suffering to Collective Experience}

\subsubsection{Symptoms and Personal Trajectories}

This theme centered on symptom documentation and illness progression. Multiple topics captured different symptom dimensions: One topic focused on brain and neural injury with keywords like `braininjury', `cognitive', and `concussion'; another addressed sensory effects including `parosmia' and `anosmia'; the next centered on fatigue and its connection to CFS; another examined respiratory symptoms; and finally one topic focused on documenting `tinnitus' and hearing loss. Patients shared detailed illness diaries, tracking symptoms like chronic fatigue, dyspnea, brain fog, loss of taste and smell, and post-exertional malaise. Daily diaries captured immediate experiences: ``Day \#XX: Today I felt winded after going from my bed to the bathroom. Went right back to bed and slept for three more hours.'' Others documented longer trajectories: ``It's been more than 10 months since I felt right...today I found that my lungs are scarred.'' This collective symptom documentation served multiple functions: validating individual experiences, establishing symptom patterns, and providing evidence for medical professionals.

As Turner et al. \citeyearpar{turner2023longcovid} document, this social consensus on symptoms was revolutionary for traditional evidence-based medicine. While individual symptoms varied and fluctuated, the collective pattern demonstrated consistent multisystem involvement. Patients were effectively crowdsourcing clinical evidence through shared experiential knowledge.

Long COVID patients explicitly encouraged documentation: ``Share details of your symptoms and trajectory—what we learn today from long COVID patients can be someone else's survival guide.'' This framing positioned patient knowledge as valuable medical data, directly challenging hierarchies that privilege clinical observation over lived experience.

\subsubsection{Epistemic Injustice, Resource Sharing, and Community Building}

Throughout symptom discussions, patients described facing medical dismissal. Many reported physicians who did not take symptoms seriously, relied exclusively on normal laboratory tests to deny illness, or suggested symptoms were psychological. Zavestoski et al. \citeyearpar{zavestoski2004patient} found that Gulf War illness patients faced similar struggles, often receiving psychiatric diagnoses (depression, anxiety, somatoform disorders) that dismissed their physical suffering. Like Gulf War veterans, long COVID patients confronted the stress of repeatedly proving illness legitimacy to skeptical clinicians, which exacerbated symptoms through anxiety \citep{zavestoski2004patient}. Clarke and James \citeyearpar{clarke2003radicalized} found that CFS patients, lacking medical validation, experienced ``anomie of suffering from a condition whose very reality is debated" in both medical and wider communities. This creates profound impacts on identity and self-concept for those with contested illnesses.

One user shared: ``My doctor told me my labs were fine, so I must be imagining things. But I can't walk to my bathroom without needing to rest for hours.'' This exemplifies how biomedical frameworks that require measurable abnormalities systematically exclude conditions with normal lab values but severe functional impairment \citep{nettleton2006just, jutel2009sociology}. Åsbring and Närvänen \citeyearpar{aasbring2003ideal} examined this dynamic from physicians' perspectives, finding that doctors treating CFS and fibromyalgia experienced frustration and professional dilemmas because these conditions could not be verified by objective measurable values. This structural tension—between medical professional ideals and the reality of treating contested conditions—shapes patient-physician encounters and contributes to dismissive care \citep{aasbring2003ideal}.

The gendered dimension of medical dismissal appeared frequently. Female patients particularly reported being told symptoms were anxiety or depression, echoing historical patterns of women's illness being attributed to hysteria or emotional instability \citep{goudsmit2009learning, khan2024psychological}.

\subsubsection{Connecting to ME/CFS and Other Contested Illnesses}

This theme shows how users explicitly linked long COVID to ME/CFS, POTS, fibromyalgia, Lyme disease, and other contested illnesses.  This connection served strategic purposes: ME/CFS communities provided templates for organizing, advocacy strategies, and warnings about harmful treatments like graded exercise therapy (GET) \citep{nezamdoust2025contested, vink2022updated}.

Twitter's affordances proved crucial for accessing these pre-existing epistemic communities. Haas \citeyearpar{haas1992introduction} defines epistemic communities as networks of professionals sharing normative beliefs, causal understandings, notions of validity, and common policy enterprises. While Haas focused on expert communities influencing international policy, we observed patient communities functioning similarly---sharing causal beliefs about post-viral illness, validating experiential knowledge, and pursuing recognition as their common enterprise. Twitter enabled long COVID patients to rapidly connect with ME/CFS advocates who had spent decades developing these shared frameworks, effectively inheriting an epistemic infrastructure that would otherwise take years to build.

Brown et al. \citeyearpar{brown2004embodied} describe ``social movement spillover''---how knowledge, networks, and strategies transfer across health movements. Twitter amplified this spillover by collapsing geographic and temporal barriers. Long COVID patients in the US could learn from UK-based ME/CFS activists' experiences with the British National Institute for Health and Care Excellence guidelines GET controversy \citep{vink2022updated}; Australian patients could access American disability advocacy strategies. This cross-national knowledge transfer, typically requiring years of conference attendance and organizational networking, occurred within months through Twitter's hashtag-mediated connections.

ME/CFS patients shared hard-won knowledge about pacing (managing energy expenditure) rather than pushing through fatigue. They warned against the dangerous recommendation of ``just exercise more,'' which can trigger severe relapses in post-viral conditions. This experiential knowledge transfer accelerated long COVID patients' understanding of their condition.

However, these connections also highlighted tensions. Some ME/CFS patients expressed frustration that long COVID received rapid research funding and attention while ME/CFS remained neglected despite decades of patient advocacy \citep{nezamdoust2025contested}. Others saw long COVID as an opportunity to finally advance understanding of post-viral illness more broadly.

Tweets detailed shared diagnostic challenges: absence of standardized tests, clinical guidelines, and institutional neglect by the CDC, NIH, and NHS. Some noted that the CDC had historically framed Chronic Lyme patients as ``anxious people whose alarm levels doctors had to manage''---making diagnosis itself a battle. Advocacy organizations bridged communities; MEAction Network (@MEActNet) used their \#MillionsMissing campaign to advocate for both ME and long COVID research, arguing that ``the pandemic has exposed...what our communities have known for decades---that government, healthcare, and society have systematically de-prioritized people with disabling complex chronic diseases.''

\subsection{Policy Advocacy and Public Health Framing}

The themes discussed next focused on COVID policy advocacy. This theme represents a shift from personal suffering to collective political action, characteristic of health social movements \citep{brown2004embodied, klein_childhood_2010}.

\subsubsection{Reframing Personal Responsibility}

Public health movements around tobacco control and obesity have successfully reframed issues from personal responsibility to structural policy \citep{kersh_politics_2009, klein_childhood_2010}. Instead of focusing on individual smoking choices, tobacco control advocates highlighted industry marketing tactics. Similarly, long COVID advocates challenged ``personal risk management" narratives. This is summarized in the message that ``they tell us to `manage our own risk,' but how do we manage risk when they won't implement basic mitigations? \#LongCOVID is a mass disabling event, not a personal choice.'' This discourse contested individualistic framings of pandemic response, arguing that population-level protections (masking mandates, ventilation improvements, paid sick leave) were necessary to prevent long COVID.

Tweets emphasized that ``COVID is not the flu'' and that post-infection effects remained poorly understood. Users warned that each new infection---even mild cases in vaccinated individuals---risked creating new long COVID patients. Urgency peaked around infection waves like Omicron, with users critiquing ``let it rip'' policies and the ``if I get it, I get it'' mentality as reckless given long COVID's severity.

\subsubsection{Schools and Children}

Connecting public health policy advocacy as opposed to private responsibility, this theme focused specifically on protecting children from long COVID. Organizations like @LongCovidKids advocated for school masking policies and proper ventilation. These campaigns drew on established frameworks from movements like Mothers Against Drunk Driving \citep{MADD25years, greenberg2021public}, positioning parents as protecting vulnerable populations from preventable harm---specifically, the uncontrolled spread of COVID in schools, which in turn would lead to long COVID in a vulnerable portion of the population.

Parents described specific symptoms in children---brain fog, fatigue---and the difficulty obtaining diagnoses. Many reported medical gaslighting when professionals unfamiliar with long COVID denied their children had significant medical issues despite detailed records documenting behavioral changes post-infection.

\subsubsection{Global Policy Variation}

Multiple topics captured international policy discourse—UK governance, Swedish COVID policy, Canadian politics, and Australian reopening policy after lockdowns---demonstrating how long COVID policy was contested across national contexts. This international coordination enabled comparative framing, with most tweets critiquing Sweden's lockdown-free approach by noting unfavorable death and infection rates compared to Nordic neighbors, while emphasizing that the pandemic was ``not over" despite political rhetoric.

This cross-national policy learning exemplifies how Twitter functions as what Haas \citeyearpar{haas1992introduction} terms a ``diffusion network" for epistemic communities. Patients observed real-time natural experiments—comparing outcomes across mask-mandating versus ``let it rip" jurisdictions, tracking long COVID prevalence under different mitigation strategies—creating a form of lay epidemiology that challenged official narratives. Brown et al. \citeyearpar{brown2004embodied} note that embodied health movements ``blur the boundary between experts and lay people" by arming activists with knowledge; Twitter accelerated this boundary-blurring by providing immediate access to international evidence deployable in local advocacy.

\subsubsection{Disability Benefits and Recognition}
This theme foregrounded disability in long COVID discourse, with many tweets ending in \#covidmaims. New rehabilitation techniques were shared, including singing and yoga for shortness of breath. Users discussed potential labor market effects as young people faced persistent symptoms in what some called a ``mass disability event,'' citing Brookings Institute estimates that long COVID removed four million Americans from the workforce \citep{bach2022longcovid}. Many noted that neither the US nor the UK were prepared for the resulting disabled population, and that was reflected in the dearth of long COVID clinics, especially in rural areas.

This theme revealed struggles accessing Social Security Disability Insurance (SSDI). Patients described being denied benefits because long COVID was not recognized as a disability category or because they could not document their initial infection (many were infected before testing was widely available). This exclusion from disability systems mirrors ME/CFS patients' experiences \citep{nezamdoust2025contested}.

The CDC estimates that 90\% of ME/CFS cases remain undiagnosed \citep{cdc_long_covid_2025}, reflecting medical systems' inability to accommodate contested illnesses. Long COVID patients faced similar challenges: without clear diagnostic criteria or biomarkers, proving disability became a kafkaesque battle with bureaucratic systems designed around visible, measurable impairments. Many reported that insurance would not cover rehabilitation or that specialists dismissed their symptoms entirely. In effect, patients noted that healthcare systems are structured for acute conditions, not chronic disease management.

This reflects broader structural issues in how medical systems handle contested illness. As Jutel \citeyearpar{jutel2009sociology, jutel2010medically} argues, without formal diagnosis, patients lose not only treatment access but social validation of their suffering. Long COVID patients found themselves in this liminal space: clearly impaired but often undiagnosed and therefore unsupported.

Discourse also addressed workplace accommodations, highlighting that many employers did not understand long COVID or refused flexibility. Like Nezamdoust and Ruel \citeyearpar{nezamdoust2025contested} documented for severe ME/CFS, patients often became too impaired to work but were denied the sick role because their condition lacked medical legitimacy.

\subsection{Building an Epistemic Community: Roles and Knowledge Networks}

Our network analysis identified distinct roles within the long COVID community that collectively constituted an epistemic community. Following Haas's \citeyearpar{haas1992introduction} definition, epistemic communities share policy orientations, causal beliefs, and notions of validity while working to shape policy. The long COVID network exhibited these characteristics while developing new forms of collective expertise.

\subsubsection{Patient Advocates and Lived Experience Experts}

This largest category included individual patients sharing their illness journeys, patient advocacy organizations (like @LongCovidKids, @LongCovidSOS), and patient-led research initiatives. These actors positioned lived experience as valid evidence, directly challenging medical hierarchies that privilege clinical observation \citep{epstein1996impure}.

Patient advocates performed crucial epistemic work: documenting symptoms, tracking rehabilitation attempts, and sharing coping strategies. As Figueiredo et al. \citeyearpar{figueiredo2024journey} demonstrate, this peer knowledge exchange supported other patients' agency in navigating their conditions.

\subsubsection{Medical Professionals and Researchers}

Healthcare providers and researchers who took long COVID seriously played bridging roles, translating between medical and patient communities. @PutrinoLab and @LongCOVIDPhysio exemplify how medical professionals could validate patient experiences while conducting formal research. This parallels how the Shapiros collaborated with Tourette Syndrome Association families to shift medical understanding of the condition \citep{kushner1999cursing}.

These alliances were strategic for both sides. Medical professionals gained access to patient populations and real-world insights, while patients gained scientific credibility. However, power asymmetries persisted: professionals could choose whether to listen to patients, while patients depended on professional validation for legitimacy.

\subsubsection{Research Coordinators and Citizen Scientists}

A surprising category emerged of patients who recruited others for research studies, shared data through platforms like Kindred where long COVID patients can share their personal journey, and even conducted independent analyses. One user celebrated: ``A prominent epidemiologist from an elite institution is now following me after I enrolled in their study! This is how we build knowledge together.'' This enthusiasm reflects both the importance of research participation and patients' exclusion from traditional knowledge production.

Citizen science initiatives, like the Long COVID Survival Guide, created patient-to-patient knowledge resources outside medical gatekeeping. This democratization of knowledge production challenges medical authority while creating alternative expertise systems \citep{epstein1996impure}.

\subsubsection{Community Organizers and Information Hubs}

Users who curated research summaries, shared policy updates, and connected patients to resources performed essential organizing labor. The hashtag \#FBLC (Follow Back Long COVID) attempted to create a distinct Twitter community, with 15 of the top 79 individual accounts using this badge.

Community building effort included new patients to ``follow [this account] if you want to know about viral infections and sequelae.'' Conversely, users warned against following medical professionals and journalists (like Nate Silver, David Leonhardt, and Leana Wen) who they saw as ``gaslighting'' patients by downplaying long COVID. This curation of trusted versus suspect sources performed boundary work, defining who belonged in the epistemic community.

\subsection{Network Structure and Tie Formation}

ERGM analysis revealed the structural characteristics driving tie formation in the long COVID network (Table \ref{tab:ergm}). The strong positive mutuality parameter shows high reciprocity---when one user mentions another, they are likely to be mentioned back, reflecting cooperative engagement.

Verified accounts were less likely to form ties (both sending and receiving mentions), suggesting the network operated outside elite Twitter structures. Sentiment analysis revealed that neutral-toned tweets were more likely to generate connections, while both positive and negative sentiment decreased tie likelihood---indicating that measured, informational discourse drove network cohesion rather than emotional content.

Crucially, users discussing ``Epistemic Injustice, Resource Sharing, \\ \& Community Building'' were significantly more likely to form ties, while those focused on ``Global Policy Variation'' were less likely. This demonstrates that the network's connective structure centered on epistemic practices---knowledge sharing and community building---rather than policy debates, supporting our characterization of long COVID Twitter as an epistemic community \citep{haas1992introduction}.

\begin{table*}[ht]
\centering
\caption{ERGM Results for Mentions Network (Significant Variables Only). *** $p<0.0001$; ** $p<0.01$}
\label{tab:ergm}
\small
\begin{tabular}{|l|c|c|c|c|}
\hline
\textbf{Variable} & \textbf{Estimate} & \textbf{Std. Error} & \textbf{z value} & \textbf{Sig.} \\
\hline
Edges (intercept) & -8.769 & 0.008 & -1059.506 & *** \\
Mutual & 5.950 & 0.037 & 160.591 & *** \\
Verified (out) & -0.116 & 0.028 & -4.119 & *** \\
Verified (in) & -0.199 & 0.029 & -6.876 & *** \\
Negative sentiment & -0.068 & 0.017 & -3.963 & *** \\
Neutral sentiment & 0.283 & 0.031 & 9.258 & *** \\
Positive sentiment & -0.157 & 0.015 & -10.627 & *** \\
\makecell[l]{Theme: Epistemic Injustice, Resource Sharing, \\ \& Community Building} & 0.181 & 0.014 & 13.029 & *** \\
Theme: Global Policy Variation & -0.206 & 0.073 & -2.837 & ** \\
Theme: Symptoms and Personal Trajectories & -0.497 & 0.063 & -7.841 & *** \\
\hline
\end{tabular}
\end{table*}

\subsection{Achieving Recognition: From Twitter to WHO}

The long COVID movement achieved formal recognition with remarkable speed. Turner et al. \citeyearpar{turner2023longcovid} document how WHO recognized long COVID by August 2020, just three months after Perego's initial tweet. This contrasts sharply with ME/CFS, which has struggled for decades to achieve legitimacy despite similar symptom profiles \citep{nezamdoust2025contested}.

\begin{figure*}
    \centering
    \includegraphics[width=1\linewidth]{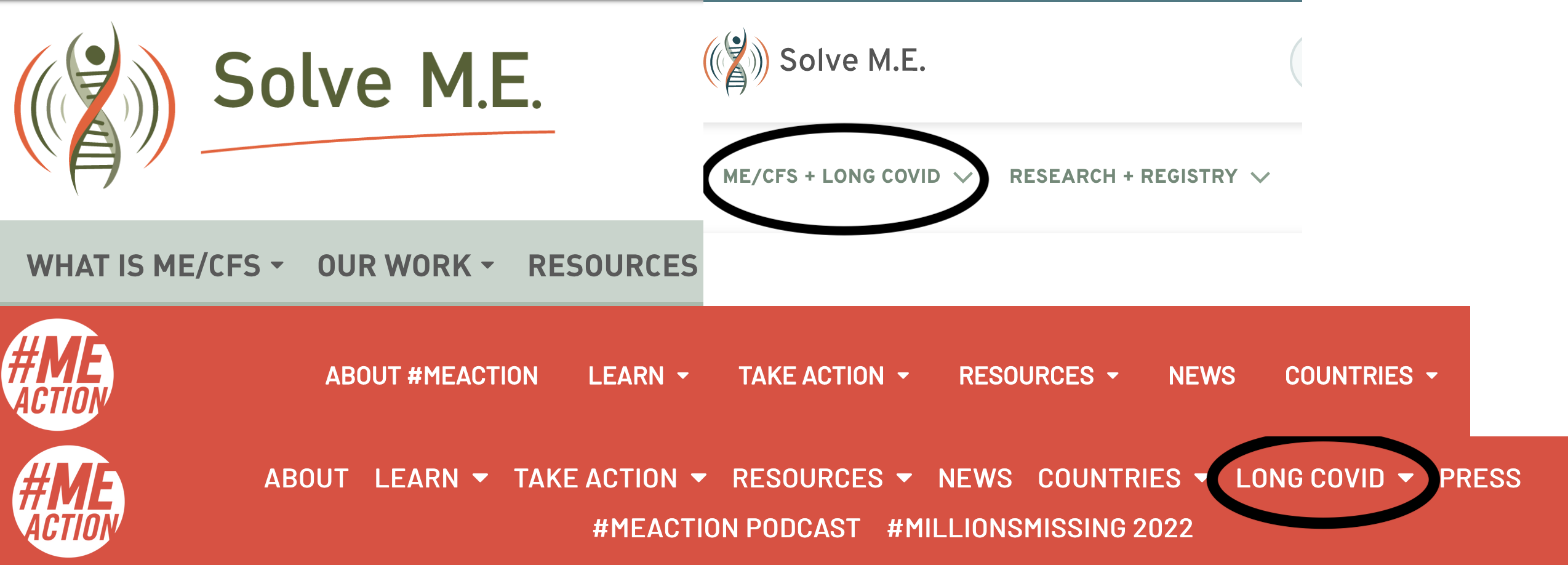}
    \caption{This figure includes four screenshots from two ME advocacy groups. For each group, a screenshot was
taken from the Wayback machine before May 2020 (first discussions of Long COVID). The other screenshot is
current. Both organizations added Long COVID to their main pages, showing the salience of the new disease
for the ME advocacy community}
    \label{fig:beforenow}
\end{figure*}

Several factors enabled this rapid recognition:

\begin{enumerate}
\item \textbf{Pandemic visibility:} Long COVID emerged during intense public focus on COVID-19, with established research infrastructure and funding.
\item \textbf{Scale:} Millions of people developed long COVID, creating a patient population too large to ignore.
\item \textbf{Social media amplification:} Twitter enabled rapid community formation and public advocacy campaigns that generated media attention.
\item \textbf{Cross-class impact:} Unlike ME/CFS, which disproportionately affects working-class women \citep{barker2011listening}, long COVID affected professionals, including healthcare workers and journalists, who commanded greater social credibility.
\item \textbf{Standing on ME/CFS shoulders:} The long COVID movement built on decades of ME/CFS patient activism while leveraging new digital tools. Long COVID advocates could draw on decades of ME/CFS organizing, avoiding pitfalls and adopting proven strategies. Two of the top organizational accounts are not long COVID advocacy organizations. They traditionally focus on advocacy for ME/CFS. Something of note is the change in the advocacy messages from ME organizations. SolveME (@PlzSolveCFS) and MEAction(@MEActNet), two ME/CFS activist groups added Long COVID to
their advocacy focus on their landing pages. In fact Solve ME literally added a “+” sign to ME on their website, making their advocacy work in ME + Long COVID. I used the Wayback Machine to find earlier versions of the organization’s websites. This is presented in Figure \ref{fig:beforenow}.
\end{enumerate}

However, formal recognition did not translate automatically into adequate care or research funding. As with ME/CFS, long COVID patients continue struggling to access disability benefits, find knowledgeable clinicians, and receive appropriate treatment \citep{figueiredo2024journey, nezamdoust2025contested}. Recognition proved necessary but insufficient for addressing the full scope of epistemic injustice.

\section{Discussion}

\subsection{Epistemic Injustice in Contested Illness}

Our findings illuminate three interconnected forms of epistemic injustice. Testimonial injustice manifested as medical professionals systematically discredited patient symptom reports, particularly for female patients---mirroring broader patterns where women's pain is undertreated and attributed to psychological causes \citep{hoffmann2001girl, khan2024psychological}. Hermeneutical injustice appeared initially when patients lacked shared language to describe their experiences; the creation of ``long COVID'' itself represented an act of hermeneutical justice, giving patients conceptual tools to articulate suffering \citep{fricker_epistemic_2007, callard_how_2021}. Institutional injustice emerged through medical systems, disability benefits, and insurance structures designed around acute conditions with clear biomarkers, excluding long COVID patients from support despite severe impairment \citep{nezamdoust2025contested}.

As Dumit \citeyearpar{dumit2006illnesses} argues, contested illnesses are conditions ``you have to fight to get,'' where diagnostic legitimacy becomes a battleground shaped by gendered and class hierarchies. Clarke and James \citeyearpar{clarke2003radicalized} show that lacking uncontested diagnosis transforms patient identity fundamentally, producing ``radicalized selves''---alternative identities forged through resistance to dismissive medical discourse and dominant cultural narratives about illness. This identity transformation reflects how contested diagnosis impacts not just access to care but self-concept, social relationships, and one's very understanding of reality when medical authorities deny the validity of one's embodied experience. CFS patients in their study experienced what they termed ``anomie of suffering from a condition whose very reality is debated'' in both medical and wider communities---a profound existential disruption that long COVID patients similarly navigate.

These injustices are not incidental to long COVID but constitutive of how contested illnesses are experienced. The pattern of gendered medical marginalization appears across contested conditions: female patients particularly reported being told symptoms were anxiety or depression, echoing historical patterns of women's illness being attributed to hysteria or emotional instability \citep{goudsmit2009learning, khan2024psychological}. Zavestoski et al. \citeyearpar{zavestoski2004patient} found that Gulf War illness patients faced similar struggles, often receiving psychiatric diagnoses that dismissed their physical suffering while the stress of repeatedly proving illness legitimacy exacerbated symptoms.

\subsection{Long COVID as an Embodied Health Movement}

Long COVID exemplifies what Brown et al. \citeyearpar{brown2004embodied} term an embodied health movement (EHM), characterized by centering bodily experience, challenging existing medical/scientific knowledge, and involving activists collaborating with sympathetic professionals. Long COVID meets all three criteria while adding a fourth: the centrality of digital platforms in rapidly mobilizing and achieving recognition.

EHMs function as ``boundary movements'' that blur distinctions between expert and lay knowledge \citep{brown2004embodied}. Long COVID activists engaged in boundary work in multiple ways. First, they challenged what counts as valid medical evidence by demonstrating that patient-generated symptom documentation could rival clinical studies in scope \citep{turner2023longcovid}. Second, they created boundary objects---shared artifacts like symptom tracking apps, the Long COVID Survival Guide, and research recruitment platforms that served both scientific and community organizing purposes. Third, some activists functioned as lay experts, using technical skills in data science and epidemiology to analyze long COVID data and build systems supporting medical professionals—paralleling how ACT UP members developed epidemiological expertise to engage with AIDS research \citep{epstein1996impure}.

Twitter's public, hashtag-based structure provided unique affordances for this boundary work. Unlike private Facebook support groups, Twitter discourse was visible to journalists, policymakers, and the general public, creating pressure on medical institutions \citep{turner2023longcovid}. The platform enabled rapid collective identity formation: rather than isolated individuals struggling to make sense of confusing symptoms, patients could immediately find community and validation. Hashtags like \#researchrehabrecognition and \#longcovidkids made patient demands visible and measurable, transforming individual suffering into collective political action characteristic of health social movements \citep{brown2004embodied}.

This process facilitated what Brown et al. \citeyearpar{brown2004embodied} term ``politicized collective illness identity''---where collective illness identity links to a broader social critique viewing structural inequalities as responsible for disease causes or triggers. Long COVID patients developed oppositional consciousness that attributed their mistreatment not merely to individual physician bias but to systemic failures in how medicine handles uncertainty and values patient expertise.

The platform also enabled cross-illness solidarity, as long COVID patients connected with ME/CFS and POTS communities, learning from their experiences and building broader coalitions \citep{nezamdoust2025contested}. ME/CFS patients shared hard-won knowledge about pacing rather than pushing through fatigue, and warned against the dangerous recommendation of ``just exercise more,'' which can trigger severe relapses in post-viral conditions. This experiential knowledge transfer accelerated long COVID patients' understanding while ME/CFS advocates saw renewed hope that research into post-viral illness might finally advance. This coalition-building answers calls by D\'{e}cary et al. \citeyearpar{decary2021humility} for collaboration between ME/CFS and long COVID advocates to develop safe care standards. Indeed, \cite{altiery2021listening} argue that “long COVID can assist in understanding the struggles and [in] mitigating the injustices faced by
similar populations. It provides an opportunity for solidarity in facing stigma, marginalization, and epistemic injustice...and rectifying systemic inequalities affecting all those with chronic diseases.”

However, digital platforms have limitations. Patient agency varies significantly---those with limited health literacy, language barriers, or severe symptoms may struggle to engage in online advocacy \citep{figueiredo2024journey}. The most visible long COVID voices may not represent the most affected populations; those with resources to maintain online presence may systematically differ from those too debilitated to participate. Online visibility does not automatically translate into medical legitimacy or healthcare access---many long COVID patients remain undiagnosed and unsupported despite extensive online organizing. Digital advocacy complements but cannot replace structural reforms to medical education, disability systems, and research priorities.

\subsection{Limitations}

This study has several limitations. First, Twitter data overrepresents users with internet access, digital literacy, and English language skills; long COVID patients from marginalized communities may be underrepresented. Second, our analysis focuses on public tweets, potentially missing important organizing in private groups or other platforms. Third, reliance on a single platform is increasingly precarious in the ``post-API'' era, where access policies and data availability can change abruptly, affecting data-deficient areas and resource-limited researchers \citep{freelon2018computational}. Future work could incorporate multiple platforms and user-contributed data through structured data donations to offset platform biases and improve representativeness \citep{ohme2024digital}. Finally, the rapidly evolving nature of long COVID means our findings capture a specific period; dynamics may shift as medical understanding develops.

Topic modeling and network analysis can identify patterns but may miss nuance. Our qualitative coding of user roles provides depth, but comprehensive interviews with long COVID advocates would yield richer understanding of organizing strategies and challenges.

\section{Conclusion}

Long COVID represents both a continuation of long-standing patterns in contested illness and a potential inflection point. Like ME/CFS patients before them, long COVID sufferers face medical dismissal, gendered stigma, and structural barriers to care and recognition---yet their unprecedented use of digital platforms enabled rapid community formation and policy impact that achieved formal recognition in months rather than decades. This case illuminates how epistemic injustice operates in medicine, where women's suffering is systematically discredited and conditions without biomarkers are treated as less real, while also demonstrating how marginalized patient communities can resist through collective organizing, knowledge production, and strategic alliances with sympathetic professionals. The COVID-19 pandemic forced medicine to confront its limitations in recognizing post-viral illness, and long COVID patients leveraged this moment to demand broader reforms to how medicine handles uncertainty, values patient expertise, and allocates disability protections. Whether this leads to lasting change remains to be seen, but the long COVID movement has demonstrated that patients need not accept medical authority passively---they can build epistemic communities that challenge, contest, and reshape medical knowledge itself.



\printcredits

\bibliographystyle{cas-model2-names}

\bibliography{cas-refs}






\end{document}